\documentclass[3p,twocolumn]{elsarticle}
\usepackage{graphicx}
\usepackage{float}
\usepackage{multicol}
\usepackage{amsmath, amsthm, amssymb}
\usepackage[small]{subfigure}

\newcommand{\paren}[1]{\left( #1 \right)}

\journal{arXiv.org}

\begin{document}

\begin{frontmatter}

\title{Study of the lateral distribution functions of electron and muon bundles using Trasgo detectors}

\author[myaddress]{A. \'Alvarez-D\'iez}
\author[myaddress]{P. Cabanelas}

\author[myaddress]{Y. Fontenla\corref{correspondingauthor}}
\cortext[correspondingauthor]{Corresponding author}
\ead{yanis.fontenla@rai.usc.es}

\author[myaddress]{J.A. Garz\'on}

\address[myaddress]{Instituto Galego de F\'isica de Altas Enerx\'ias, University of Santiago de Compostela, Spain.}

\begin{abstract}
Some of the main features of the new generation Trasgo detectors are their capability in measuring the incoming direction and the arrival time of secondary cosmic particles. They also offer the identification capability between muon and electrons and a rough calorimetry for electrons. Using ground-based stations, these properties allow for the development of new tools for the measurement of primary cosmic ray fluxes. In order to verify and quantify the suitability of Trasgo detectors, whether a single one or arrays of them, to provide reliable information of the properties (mass, energy, incoming direction) of primary cosmic rays we have started an initiative for the systematic study of the 'lateral distributions' displayed by electrons and muons, or by bundles of those particles, using MonteCarlo simulations. In a first approach, electrons and muons were produced in vertical showers from primary H, He, C and Fe nuclei, and with incoming energies limited to a maximum of 10$^{15}$ eV per nucleon. This choice represents a significant component of all secondary particles, which can be measured on Earth's surface. The lateral distributions study has been done at the two locations of Santiago de Compostela (Spain) and Livingston Island (Antarctica), where Trasgo detectors are either in operation, or will be operative in the near future.
\end{abstract}

\begin{keyword}
Cosmic Rays \sep EAS \sep Trasgo detectors
\end{keyword}

\end{frontmatter}

\section{Introduction}
Nowadays, it is assumed by the community that 74\% of the primary Cosmic Rays that arrive to the Earth are protons, and approximately 18\% are Helium nuclei. Those primaries collide with the atmosphere, generating successive interactions produced by fragments of the collision and make an atmospheric Extensive Air Shower, EAS. The electromagnetic component of the shower, made by secondary electrons and muons, is studied in this work. Secondary counting at ground level is carried out with the Trasgo Family detectors \cite{trasgo}, and, in particular, with Tragaldabas \cite{tragaldabas} and Tristan \cite{tristan} detector systems. Both are cosmic ray telescope based on timing RPC technology. Tragaldabas has a height of 1.8 m and consists of 4 RPC detection planes of 2.2 m$^2$ each, segmented in 120 RPC pads per plane, and is taking data from the Faculty of Physics of the University of Santiago de Compostela, USC, Spain (42$^{\mathrm{o}}$52'N, 8$^{\mathrm{o}}$33'W) since 2015. The Tristan detector is a 48.4 cm height system, with 3 detection planes of 1.84 m$^2$ each and segmented in 30 RPC pads per plane, and was set up to collect data at the Antartic Spanish Base in Livingston Island, Antarctica (62$^{\mathrm{o}}$37'S, 60$^{\mathrm{o}}$27'W).

On the other hand, the CORSIKA event generator simulation program \cite{corsika} has been used for the generation of simulated data. A deep analysis of the components and lateral distributions of the secondary showers over the Earth's surface was done. The observed behaviour will allow us to know the performance of the detector system mentioned above under extensive air shower. We present here the simulated response of the detectors, the radial density of the showers and the main properties of the particle clusters that can be detected at ground level as a function of the nature of the primary Cosmic Ray.

\section{Generation of simulated data}
The simulations and propagations of the different EAS were performed with the CORSIKA event generator software. Four different primary nuclei, H, He, C and Fe, with an energy range from 1 GeV to 10$^5 $ GeV and a zenithal angle from 0 to 58.3$^\circ$ were launched against the Earth atmosphere and EAS were produced. The energy  range of primary cosmic rays was set in steps of 4$\times$10$^{1/8}$ per decade (logarithmic scale), while the zenithal angle distribution has a uniform separation in steps of 0.025 in $\cos(\theta)$. For each situation, they were studied EAS events generated from 10$^4$ to 10$^2$ primaries, as they increased in energy, and the first interaction of the primary nucleus happened at an altitude between 35 km and 18 km. The GHEISHA and QGSJET models were used in CORSIKA for the low energies simulations. It was set an energy cut of 0.05 GeV for hadrons and muons and 0.09 GeV for electrons and photons.

\section{Methods}

\begin{figure}[htb!]
\centering
\includegraphics[width=0.45\textwidth]{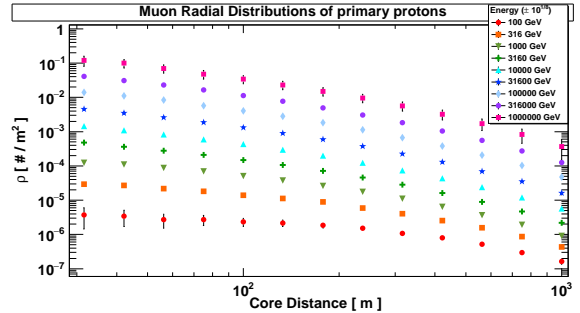}
\caption{Radial distribution of muons at the ground as a function of the core distance for primary protons with vertical incidence.}
\label{fig1}
\end{figure}

An EAS is characterized by the size and the primary energy. The size of an individual shower can be determined by sampling the particle density distribution at ground level with an array of suitable detectors over the EAS action area. The density of particles or the lateral density distribution function (LDF), $\rho$ [m$^{\text{-}2}$] can be written as:

\begin{equation}
\rho(\text{r})=\text{C}\cdot\paren{\frac{\text{r}}{\text{r}_0}}^{\alpha}\cdot\paren{1+\frac{\text{r}}{\text{r}_0}}^{\beta}
\label{ec:1}
\end{equation}

\noindent where $r$ is the distance to the geometric center of the EAS and $C$, $r_0$, $\alpha$ and $\beta$ are free parameters. T. Hara \cite{hara} used this equation to fit his experimental data and obtained the following values: $r_0$=280 m, $\alpha$=-0.75 and $\beta$=-2.5. The same equation is used also to fit the data in this work and represent the parameters as a function of the energy of the primary cosmic ray. Figure \ref{fig1} shows the lateral distributions of density muons at ground level from primary protons of different energies and for an $r$ value ranging from 31.6 m to 1000 m.

\begin{figure}[htb!]
\centering
\includegraphics[width=0.45\textwidth]{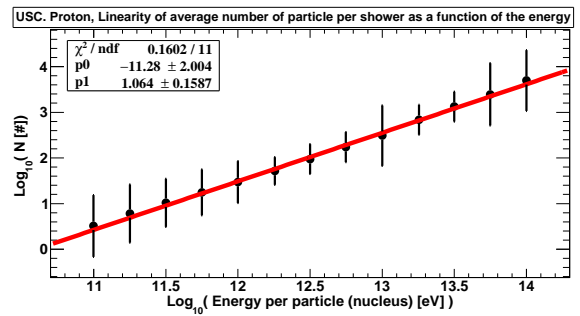}
\caption{Size of the shower as a function of the primary protons energy with vertical incident in the atmosphere.}\label{fig2}
\end{figure}

On the other hand, the size $N_0$, which is nothing but the number of particles of the shower, is proportional to the energy of the primary cosmic ray, and can be determine by:

\begin{equation}
\text{N}_0\simeq\frac{1}{\kappa^{1/b}}\cdot\text{E}^{1/b}_0 
\label{ec:2}
\end{equation} 

\noindent where the parameters $\kappa$ and $b$ take the value of $\kappa$=$2.217\cdot 10^{11}$ and $b$=0.798 for protons (A.M. Hillas \cite{hillas}). Figure \ref{fig2} shows the average particles mix of shower counted as a function of the primary proton energy with vertical incidence in the atmosphere. The data perfectly reflects a linear behaviour. The results of the fit have values of $\kappa$=4$\cdot10^{10}$ y $b$=0.94. These results are in agreement with the previous results of A.M. Hillas.

\begin{figure}[htb!]
\centering
\includegraphics[width=0.45\textwidth]{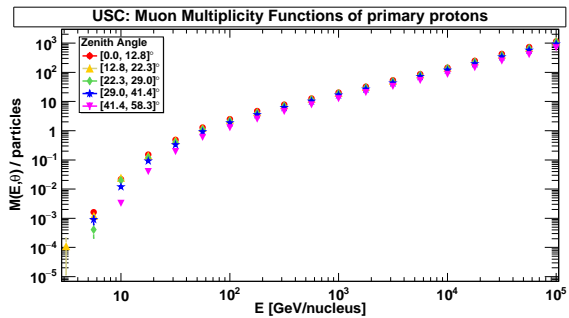}
\caption{Multiplicity function of muons for primary protons calculated for the geography location of the University of Santiago de Compostela, USC, Spain.}\label{fig3}
\end{figure}

The detector response is an important factor used to know the rate of  particles arriving at ground level. The response of the detectors is calculated with the coupling functions \cite{yakovleva}. The functions link the spectrum of cosmic ray energies with a number of particles counted at ground level. The Yield function or commonly called response, G(E,$\theta_i,\varphi_i$) [GeV$^{\text{-}1}\cdot$s$^{\text{-}1}$], is the number of muons integrated into a solid angle $S(\theta_i,\varphi_i)$ with effective area in a given direction $(\theta_i,\varphi_i)$. It can be written as:

\begin{equation}
G(E,\theta_i,\varphi_i)=M(E,\theta_i) \cdot J_p(E)\cdot \Delta\,S \Omega(\theta_i,\varphi_i)
\label{ec:3}
\end{equation}

\noindent where $M(E,\theta_i)$ is the multiplicity function, $\Delta\,S \Omega(\theta_i,\varphi_i)$ is the partial acceptance of the detector according to $ (\theta_i, \varphi_i) $ and $ J_p (E) $ is the differential energy ($ J_p (E) $ could depend on the angles $ \theta $ and $ \varphi $). This equation is used to calculate the response of both the Tragaldabas detector, at the geographic location of the USC, Spain, and for the Tristan detector, which will be located in the Antarctic Spanish Base in Livingston, Antarctica. Figure \ref{fig3} shows the muon multiplicity function of calculated at the Tragaldabas location for incident primary protons. The data were calculated for the different angles of vertical incidence of the particle in the atmosphere. The primary flux equations used to calculate the response curves were taken from the reference \cite{pdg_CR}.

\section{Results}
\begin{figure}[htb!]
\centering
\subfigure[]{\includegraphics[width=0.45\textwidth]{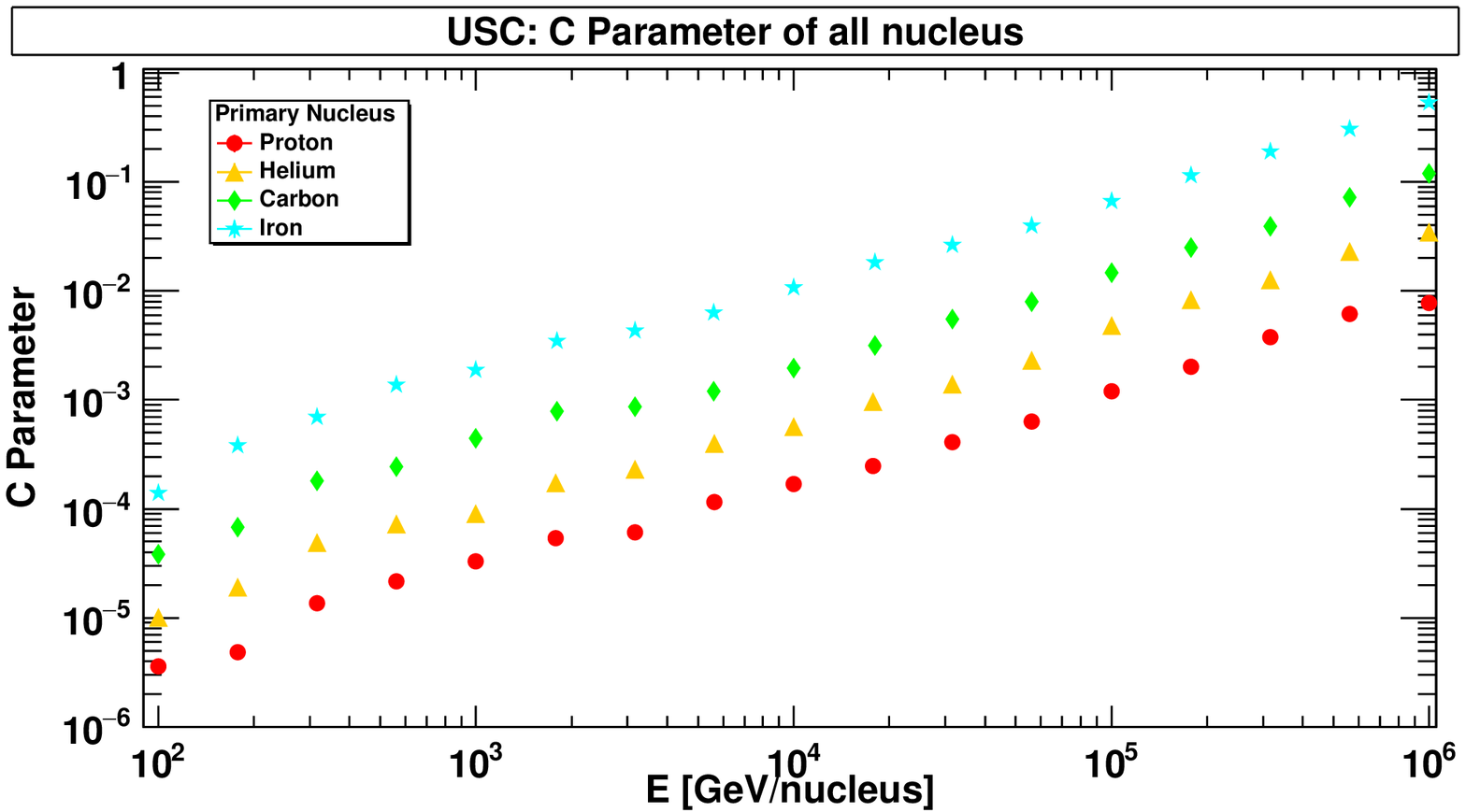}}
\subfigure[]{\includegraphics[width=0.45\textwidth]{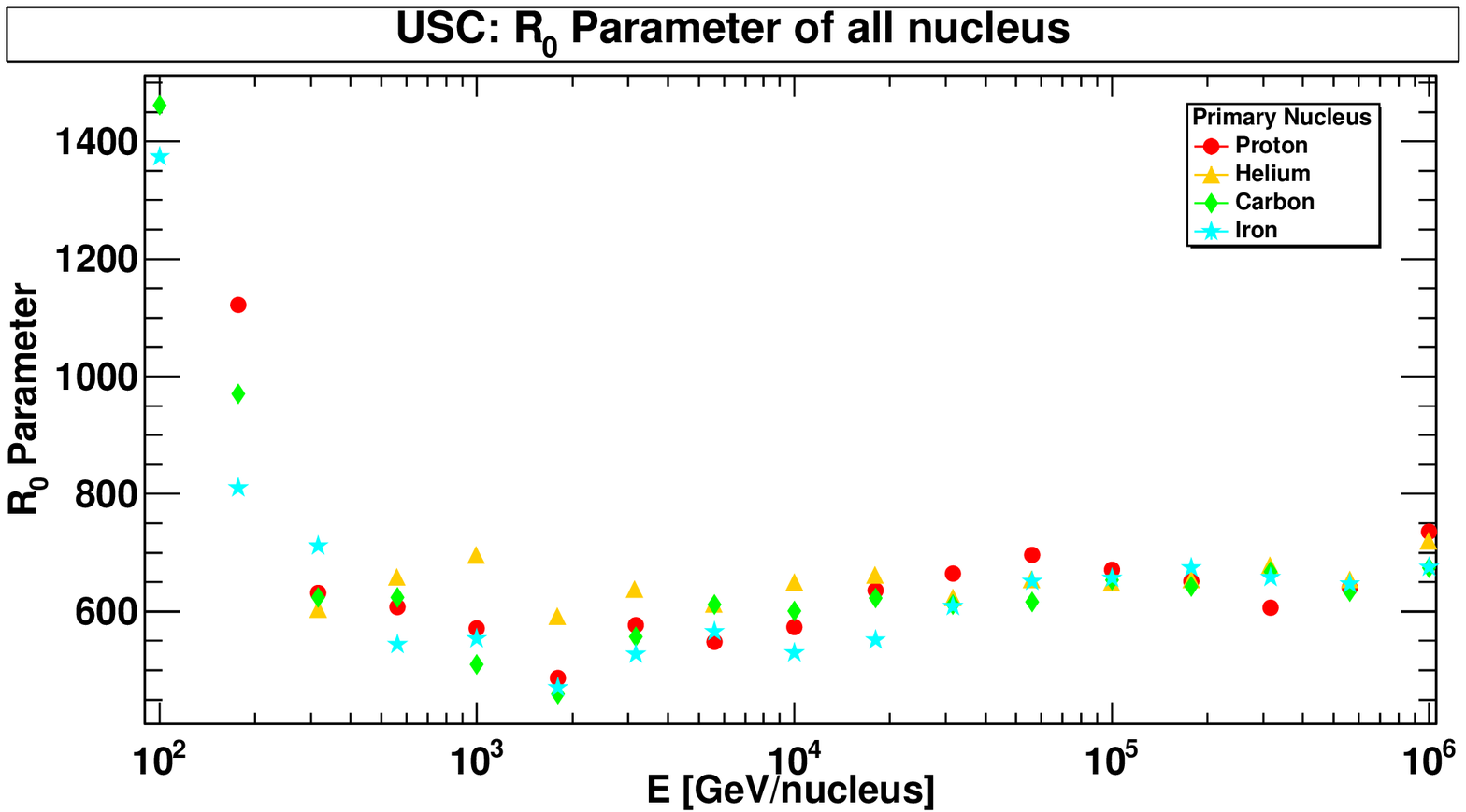}}
\caption{Parametrization of C (a) and r$_0$ (b) LDF parameter value for nuclei with vertical incidence in the atmosphere.}\label{fig4}
\end{figure}

The parameters for the LDF in equation \eqref{ec:1} were calculated by means of the data presented in Figure \ref{fig3} for a vertical incidence of primary protons at different energies. Figure \ref{fig4} (a) shows the $C$ parameter as a function of the energy for the different primary nuclei. It is observed a close to linear relationship between $C$ and the primary energy, which indicates that the $C$ parameter is also directly correlated with the number particles $N_0$.

Figure \ref{fig4} (b) presents the $r_0$ parameter as a function of energy of the primary nucleus. The results data have an asymptotic and oscillating behavior. It is observed that for energies above 10$^3$ GeV $r_0$ can be assumed as constant with slight deviations around a given value of about 650 m, but presents an exponential growth as the energy of the primary decreases. They were obtained also $\alpha$ and $\beta$ values around -0.9 and -3.7 respectively. All these results are in agreement with the experimental values obtained in \cite{hara}.
 
In order to a better study and understanding of the secondary particles, they were divided into clusters. A cluster is a close region of the ground level which contains a given number of secondary particles. The position of the center of each cluster was calculated by implementing an iterative algorithm in the CORSIKA routine, trying to maximize the number of particles per cluster. It was chosen a cluster radius of 2 m$^2$ and a cut-off standard deviation from the first radius of interaction of 0.5 m. Trying to simulate also the particle detection with the Tragaldabas detector, a cut-off energy of 100 MeV for electrons and 200 MeV for photons was implemented. The number of secondary particles inside a circular section cluster was taken as the corresponding cluster's size. 

\begin{figure}[htb!]
\centering
\subfigure[]{\includegraphics[width=0.45\textwidth]{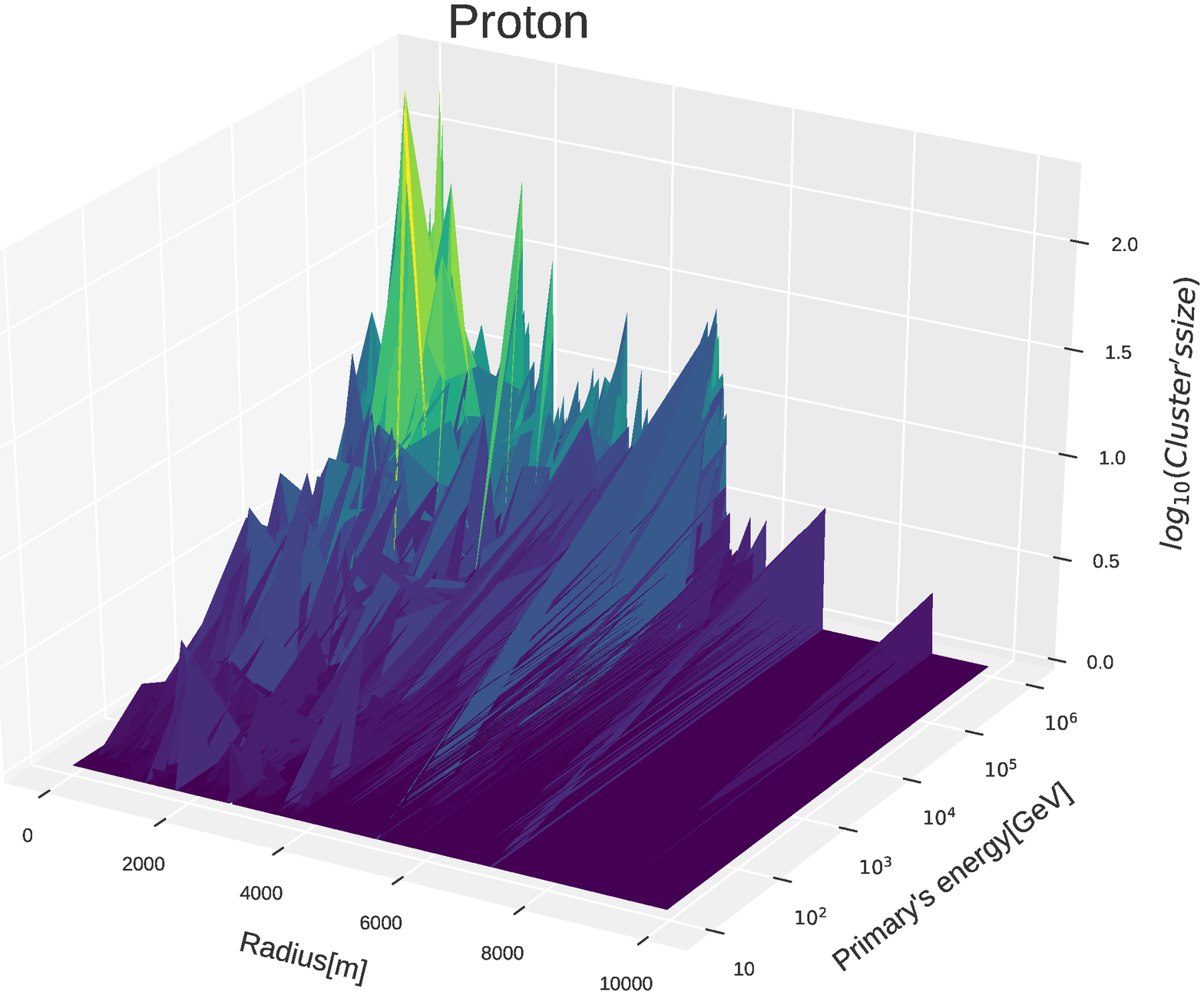}}
\subfigure[]{\includegraphics[width=0.45\textwidth]{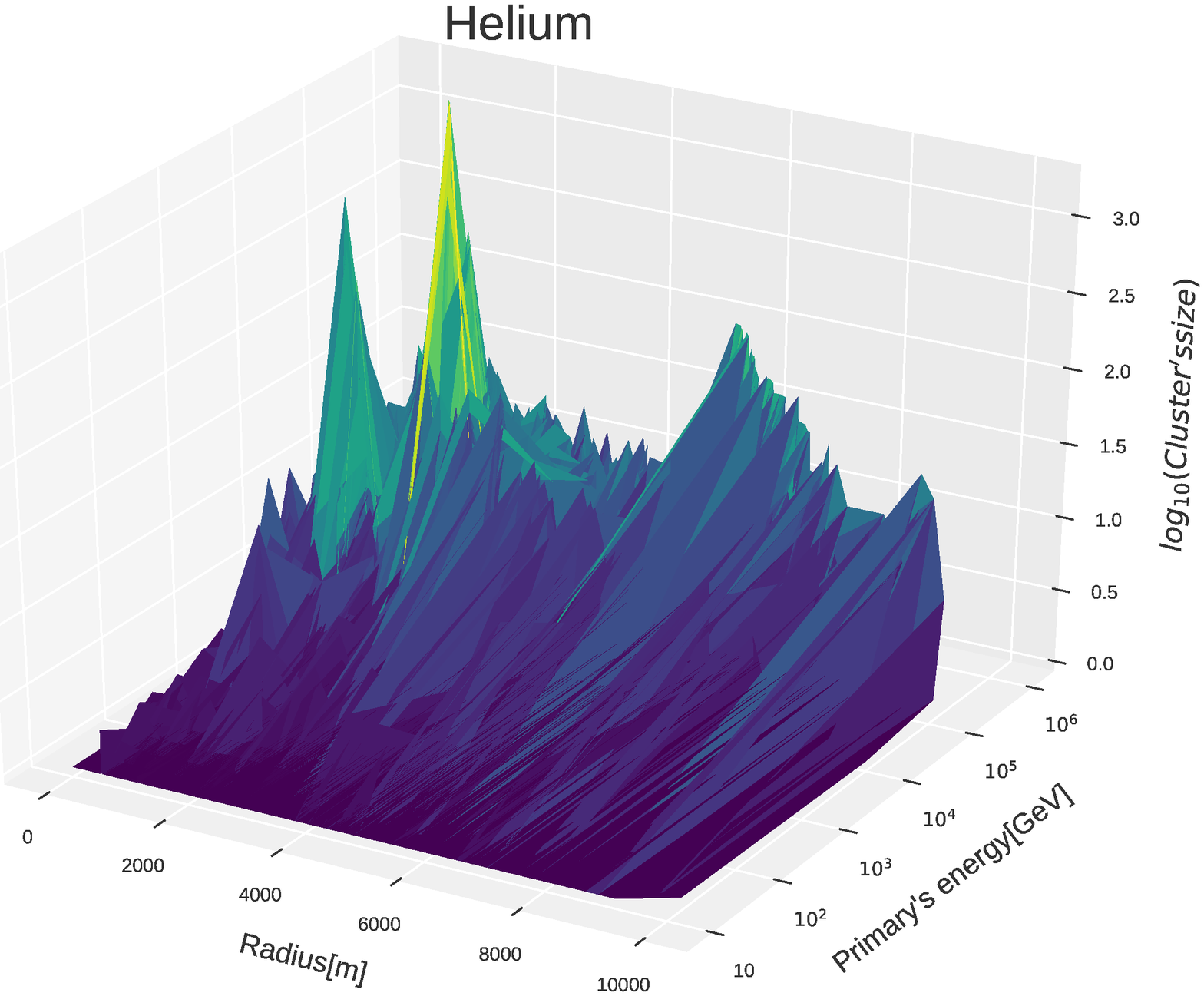}}
\caption{Cluster size representation as a function of the distance to the EAS core, $radius$, and the primary nucleus energy, both for primary protons (a) and He nuclei (b).} \label{fig6}
\end{figure}

In Figure \ref{fig6} it is shown the cluster size for vertical incidence as a function of the distance between the center of the cluster and the core of the EAS, called $radius$, and the primary energy, for two different primary nuclei. The radius and the cluster size were calculated by sorting the data by the radius and separating them by sections of 10 m. Then, for each section, it was calculated the mean cluster size and the mean radius. The data was plotted by triangulation using Python2.7. At low energies, the cluster size tends to the unity regardless of the cluster radius. When the energy grows there is a raise of the cluster size that becomes more pronounced as the radius decrease. One can also see how that behaviour is more pronounced as the primary becomes heavier. We believe that the primary source and its energy can be identified with the localized information of a variety of clusters.

The equation \eqref{ec:3} was used to calculate the response function of the detectors, $G(E,\theta_i,\varphi_i)$, at the ground level under irradiation from EAS. Results are collected in Figure \ref{fig3}, where a fit to a fifth order polynomial function was imposed. The response distributions are calculated as the product of the polynomial equation and the primary flux equations at the Earth's ground.

\begin{figure}[htb!]
\centering
\subfigure[]{\includegraphics[width=0.45\textwidth]{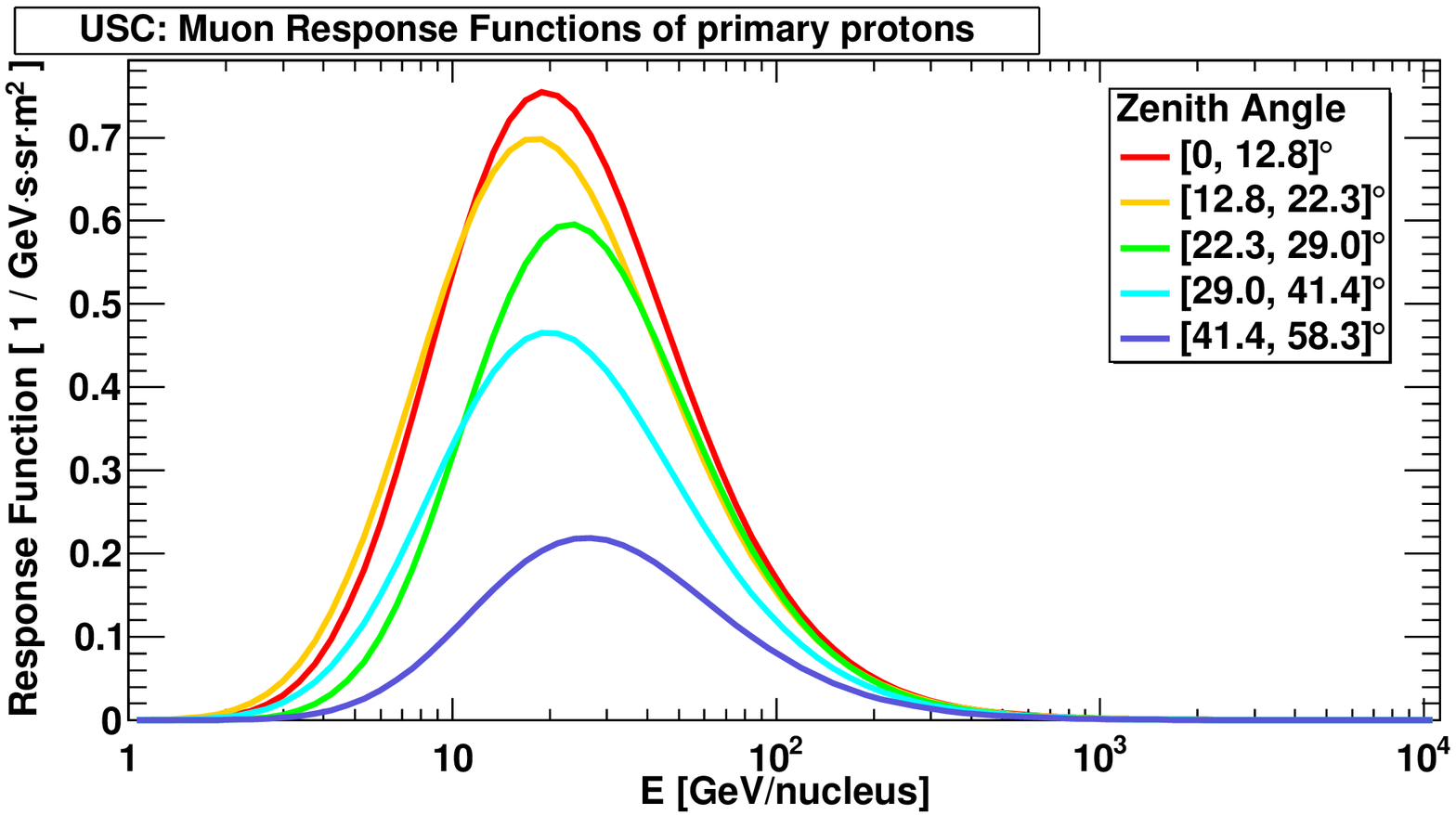}}
\subfigure[]{\includegraphics[width=0.45\textwidth]{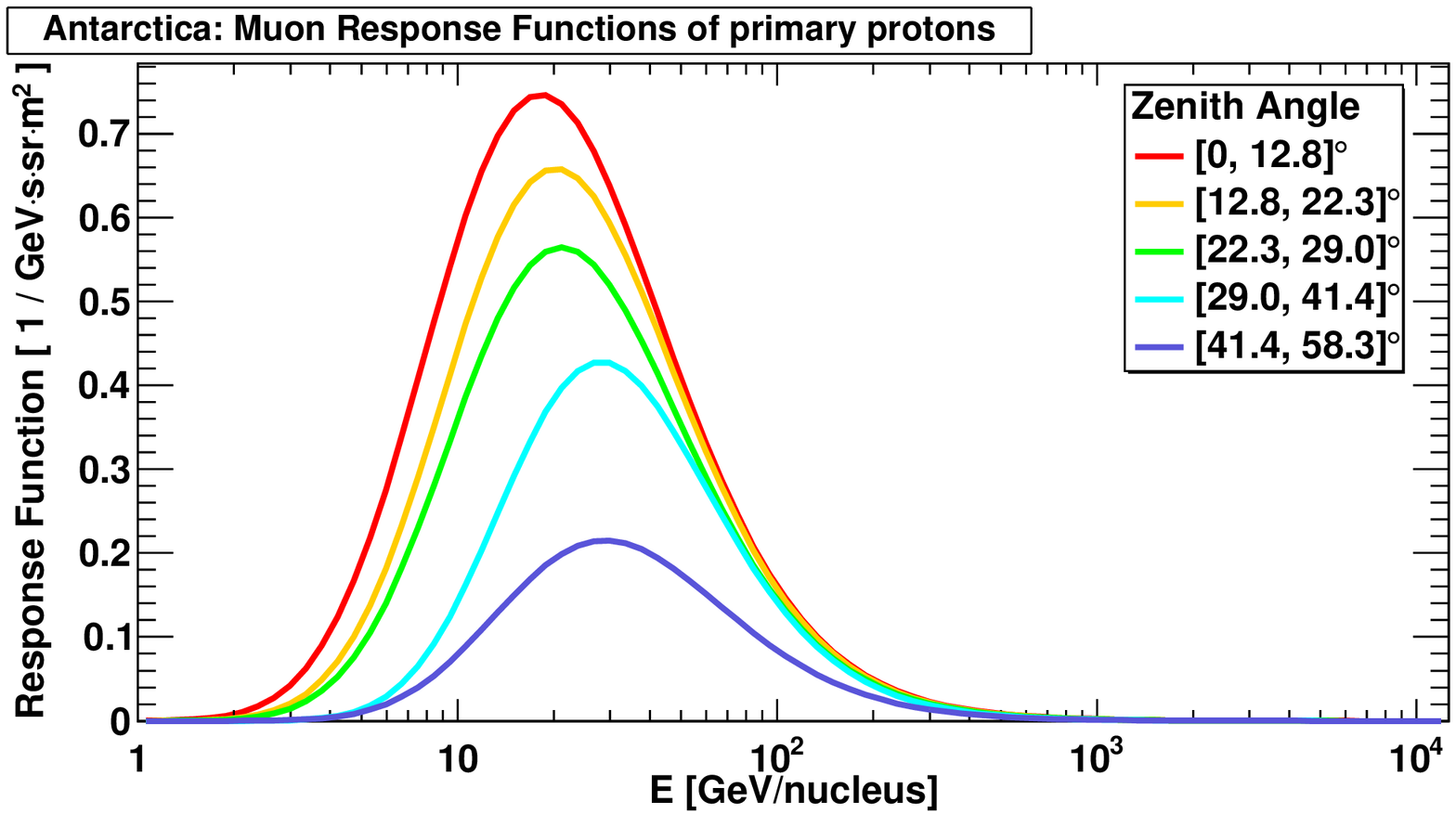}}
\caption{Response function of detected muons calculated in the geographic location of USC, Spain (a) and in the geographic location of the Antarctic Spanish Base, Antarctica (b). The secondary muons are produced by vertical incidence of protons in the atmosphere.}\label{fig5}
\end{figure}

Figure \ref{fig5} shows the response function distribution of muons at the University of Santiago de Compostela geographic location ground level for different incidence angles of primary protons in the atmosphere. The response decreases as the zenithal angle grows. Also, the mean of the distributions is shifted towards higher energies with the angle. The mean of the distribution is around 20 GeV for small angles and goes up to 30 GeV for larger angles.

The muon response is calculated also for different incident zenithal angles in the Antarctic Spanish Base location. The trend and the mean of the distributions are slightly different than the case of the USC location. Distributions now are less oscillating as the angle grows faster than in the previous case.

\section{Tragaldabas and Tristan Detectors specifications}
\begin{figure}[htb!]
\centering
\subfigure[]{\includegraphics[width=0.45\textwidth]{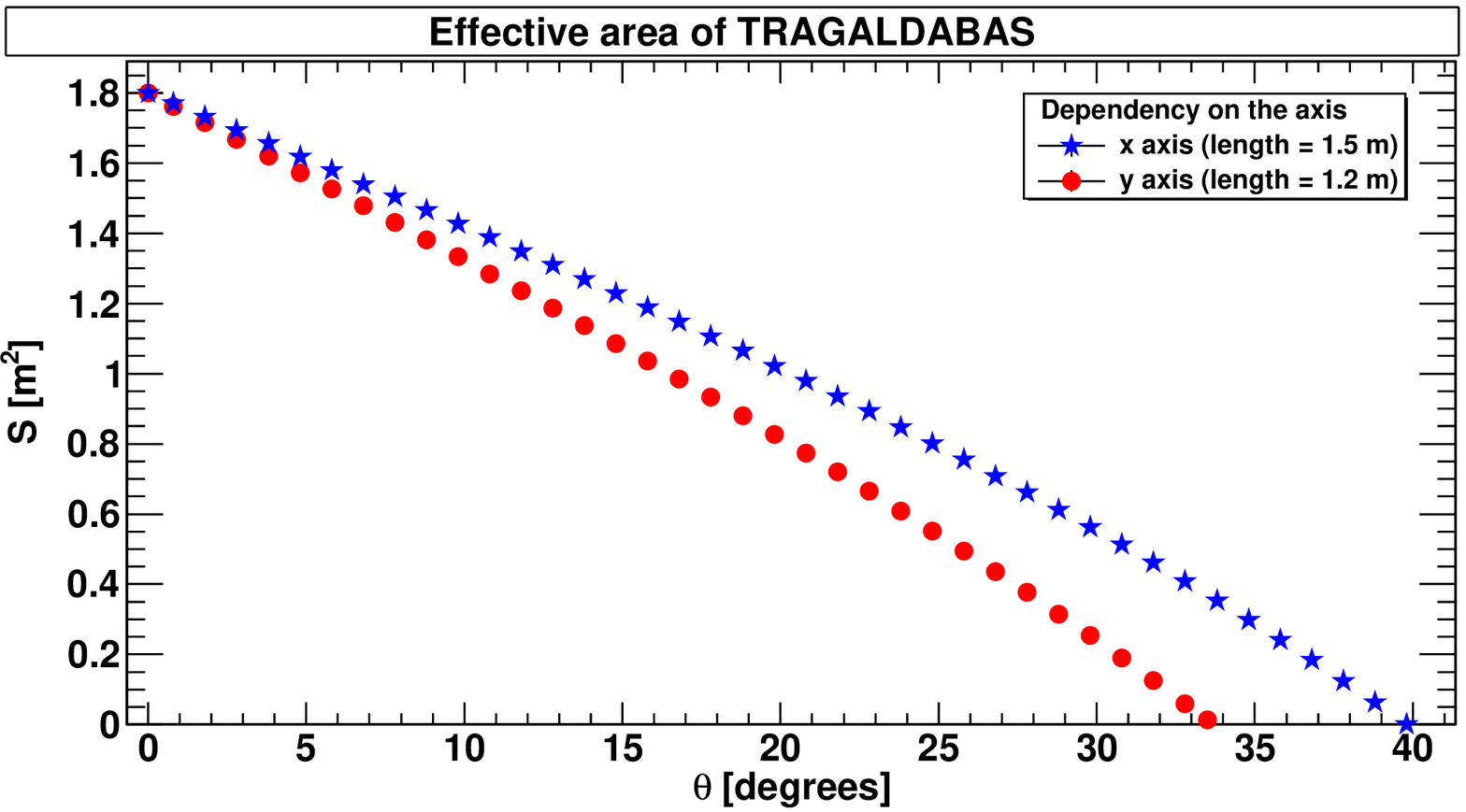}}
\subfigure[]{\includegraphics[width=0.45\textwidth]{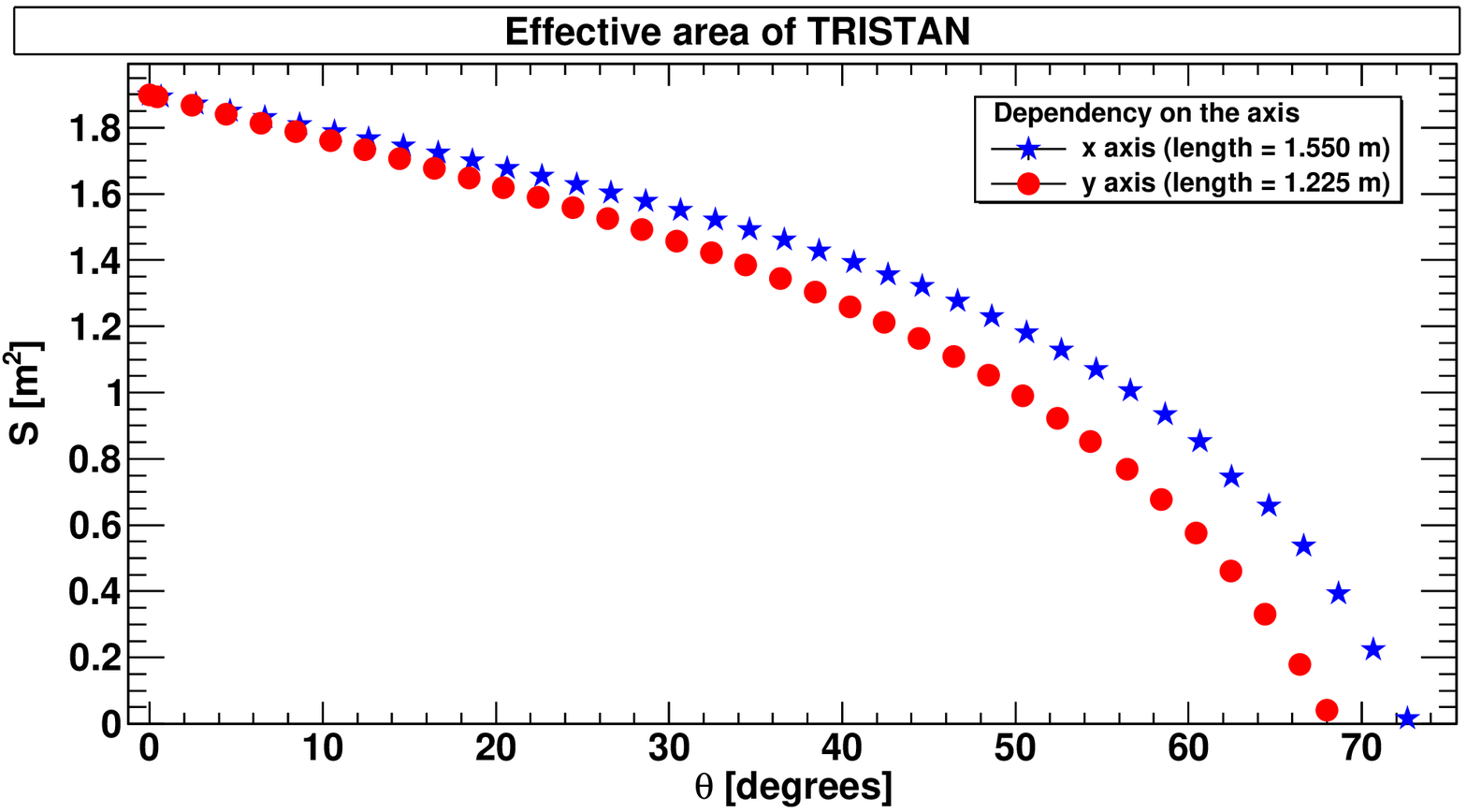}}
\caption{Calculated effective area for the both Trasgo-like systems, the Tragaldabas detector (a) and the Tristan detector (b), calculated for the $x$-axis and the $y$-axis references. The different observed for the different coordinates systems is due to the fact that both detector systems are not square shaped, but rectangular shaped.}
\label{fig7}
\end{figure}

As the cosmic rays particle rate over a Trasgo-like detector is directly correlated with the detector configuration, geometry and effective detection area, the response function distribution varies according to the set up used. The layout currently used with the Tragaldabas and Tristan detectors involves the effective area for particle detection as a function of its impinging direction presented in Figure \ref{fig7}. Given the fact that both systems have a rectangular shape, that effective area can change according to the calculations whether they are in the $x$-axis or in the $y$-axis.

Table \ref{tab1} shows the partial acceptance for the two studied Trasgo-like detector systems, Tragaldabas and Tristan. Those acceptances can be used for a later rescaling of the response function distributions calculated in the Figure \ref{fig5}. The configuration of three active planes of Tristan as compared with the configuration of four active planes of Tragaldabas makes really a big difference in the effective detection area when detecting particles arriving with high angles, being greater for the first. On the opposite,  the four planes configuration ensures more confidence about the accepted triggers.

\begin{table}[htb!] 
\centering 
{
\begin{tabular}{c|c|c} 
\cline{2-3} 
&  \multicolumn{2}{c|}{$\Delta\Omega \cdot S$ [sr$\cdot$m$^{2}$]}
\\
\hline
\multicolumn{1}{|c|}{$\theta$ Range [degrees]} & 
\multicolumn{1}{c|}{Tragaldabas} & \multicolumn{1}{c|}{Tristan}
\\
\hline 
\multicolumn{1}{|c|}{[0, 12.8]} & \multicolumn{1}{c|}{0.28} & \multicolumn{1}{c|}{0.30}\\  
\hline
\multicolumn{1}{|c|}{[12.8, 22.3]} & \multicolumn{1}{c|}{0.54} & \multicolumn{1}{c|}{0.60}\\ 
\hline
\multicolumn{1}{|c|}{[22.3, 29]} & \multicolumn{1}{c|}{0.52} & \multicolumn{1}{c|}{0.60}\\  
\hline
\multicolumn{1}{|c|}{[29, 41.4]} & \multicolumn{1}{c|}{1.15} & \multicolumn{1}{c|}{1.49}\\ 
\hline 
\multicolumn{1}{|c|}{[41.4, 58.3]} & \multicolumn{1}{c|}{1.64} & \multicolumn{1}{c|}{2.68}\\ 
\hline 
\end{tabular}
}
\caption{\small Partial acceptance, in $\Delta\Omega \cdot S$ [sr$\cdot$m$^{2}$], obtained for different ranges of zenith angles, both for the Tragaldabas and the Tristan detectors systems. The bigger acceptance of the Tristan detector at higher incident angles is due to the three detection planes configuration of the system, while a four planes configuration is used for Tragaldabas.}
\label{tab1}
\end{table}

\section{Conclusions}
A study of lateral radial distribution was carried out and implemented in this work. The shower energies and radial densities results show good agreement with the experimental results of both A.M. Milas as of T. Hara respectively. The response function curves for two of the existing Trasgo-like detector systems, the Tragaldabas detector, at University of Santiago de Compostela, Spain, and the Tristan detector, which will be located at the Antarctic Spanish Base, Livingston Island, Antartica, were calculated by means of simulation. Those response functions can give information about the properties of the incident primary cosmic ray in the atmosphere, such as the mass, the energy and the arrival direction, since the rate of particles that arrive at ground level is then well known. The 3D representations for incident H and He nuclei at the atmosphere provide us with an estimate of the primary's energy and an approximation the structure of the EAS at the ground level over a large detection area.

\section*{Acknowledgements}
This work has been financially supported by the Plan Galego de Investigaci\'on, Innovaci\'on e Crecemento (I2C) of Xunta de Galicia under project ED431C 2017/54.
Some results presented in this paper were obtained thanks to the access granted to Galicia Supercomputing Center (CESGA) high performance computing resources.


\section*{References}

\begin{thebibliography}{99}
\bibitem{trasgo} D. Belver \emph{et al.}, \emph{TRASGO: A proposal for a timing RPCs based detector for analyzing cosmic ray air showers}, Nucl.Inst.Meth. A, {\bf 661} (2012) S163-S167
\bibitem{tragaldabas} H. \'Alvarez-Pol \emph{et al.}, \emph{Tragaldabas: A new high resolution detector for the regular study of cosmic rays}, J.Phys.:Conf.Ser. {\bf 632} (2015), no. 1, 012010
\bibitem{tristan} D. Garc\'ia-Castro, P. Cabanelas, J.A. Garz\'on \emph{et al.}, \emph{The TRISTAN Antarctic Cosmic Ray detector}, PoS (ICRC2019) 071.
\bibitem{corsika} D. Heck \emph{et al.}, Report FZKA 6019 (1998). Forschungszentrum Karlsruhe. http://www-ik.fzk.de/corsika/physics description/corsika phy.html.
\bibitem{hara} T. Hara \emph{et al.}, \emph{Characteristics of large Air Showers at ore distances between 1km and 2km}, 11 (1983) 276.
\bibitem{hillas} A.M. Hillas,\emph{Some recent development in cosmic rays}. (1975) Phys. Rep. C. 20, 79
\bibitem{yakovleva} Elena I Yakovleva \emph{et al.},\emph{Coupling functions for muon hodoscopes}. (2009)  Bulletin of the Russian Academy of Sciences: Physics. 73. 10.3103/S106287380903023X
\bibitem{pdg_CR} M. Tanabashi \emph{et al.}, \emph{(Particle Data Group)}, Phys. Rev. D 98, 030001 (2018), p.424.
\end{thebibliography}
\end{document}